\documentclass[final,1p,times]{elsarticle}
\usepackage{graphicx}
\usepackage{url}

\usepackage{amssymb}
\usepackage{amsmath}
\journal{Proceedings of the Combustion Institute}

\begin{document}
%\small
%\baselineskip 10pt
\begin{frontmatter}


\title{Influence of hydrodynamic instabilities on the propagation mechanism of fast flames}

\author{L. Maley, R. Bhattacharjee, S. Lau-Chapdelaine, M.I. Radulescu, }

\address{Department of Mechanical Engineering, University of Ottawa, Ottawa Canada K1N6N5}

\begin{abstract}
The present work investigates the structure of fast supersonic turbulent flames typically observed as precursors to the onset of detonation.  These high speed deflagrations are obtained after the interaction of a detonation wave with cylindrical obstacles. Two mixtures having the same propensity for local hot spot formation were considered, namely hydrogen-oxygen and methane-oxygen.  It was shown that the methane mixture sustained turbulent fast flames, while the hydrogen mixture did not.  Detailed high speed visualizations of nearly two-dimensional flow fields permitted to identify the key mechanism involved.  The strong vorticity generation associated with shock reflections in methane permitted to drive jets.  These provided local enhancement of mixing rates, sustenance of pressure waves, organization of the front in stronger fewer modes and eventually the transition to detonation.  In the hydrogen system, for similar thermo-chemical parameters, the absence of these jets did not permit to establish such fast flames.  This jetting slip line instability in shock reflections (and lack thereof in hydrogen) was correlated with the value of the isentropic exponent and its control of Mach shock jetting instability (Mach \& Radulescu).    
\end{abstract}


\end{frontmatter}
% \linenumbers
%% main tex
\section{Introduction}
\addvspace{10pt}
The last stage of the deflagration to detonation transition is the so-called fast flame, or choking regime, where a deflagration and shock wave complex propagates quasi-steadily at approximately half the CJ velocity. This regime is presently very poorly understood, sometimes even labeled as a ``strange wave" \cite{Thomasetal2001}. The burning front observed experimentally appears as a highly turbulent reaction zone overlapping with the front shock and displaying very strong pressure fluctuations \cite{Thomasetal2001, Grondin&Lee2010, Ciccarellietal2013}. Burning velocities of a few hundred meters per second are required behind the leading shock in order to explain such waves. It is difficult to account for such high deflagration speeds based on turbulent deflagration velocities; experiments in fan-stirred bombs show that turbulent velocities eventually saturate as the turbulent intensity is increased \cite{Bradley2012}. It is also difficult to account for these waves as a diffusionless phenomenon alone, since the auto-ignition delay in the gas compressed by the shock is not compatible with the reaction wave propagation, unless strong temperature fluctuations are present to trigger local ignition kernels (hot spots) with ignition delays compatible with the observed reaction wave propagation \cite{Bhattacharjeeetal2013, Radulescu&Maxwell2011}. Based on indirect and often speculative experimental observations, it is believed that these fast deflagration velocities are maintained via the action of pressure waves within this deflagration complex (see \cite{Ciccarellietal2013} and references to earlier work by Lee and his collaborators \cite{Lee2008}). These transverse pressure waves are believed to contribute to the increase of the level of turbulent mixing by the Richtmyer-Meshkov instability, similar to the action of transverse waves in highly unstable cellular detonations. 

The experimental configuration used to isolate these high speed waves, without relying on a prior low speed flame acceleration, is generally obtained through the interaction of a detonation wave with a porous plate \cite{Chao2006, Grondin&Lee2010}, from which a transmitted shock - turbulent flame complex can emerge.  The compressible turbulent interactions driven by the shock diffraction and reflections immediately behind the porous plate provide the first strong perturbations, but these soon die out in relatively stable mixtures, who fail to support such fast flames \cite{Grondin&Lee2010}.  It was found that such fast deflagration waves are only preferentially established in unstable mixtures \cite{Grondin&Lee2010, Chao2006, Kuznetsovetal2000}, i.e., mixtures with high activation energies and high ratio of the characterictic induction to reaction times, conditions that favor the amplification of pressure disturbances from local hotspots \cite{Bradley2012, Radulescuetal2013}.  

\begin{figure}
\begin{center}
\includegraphics[width=0.8\textwidth]{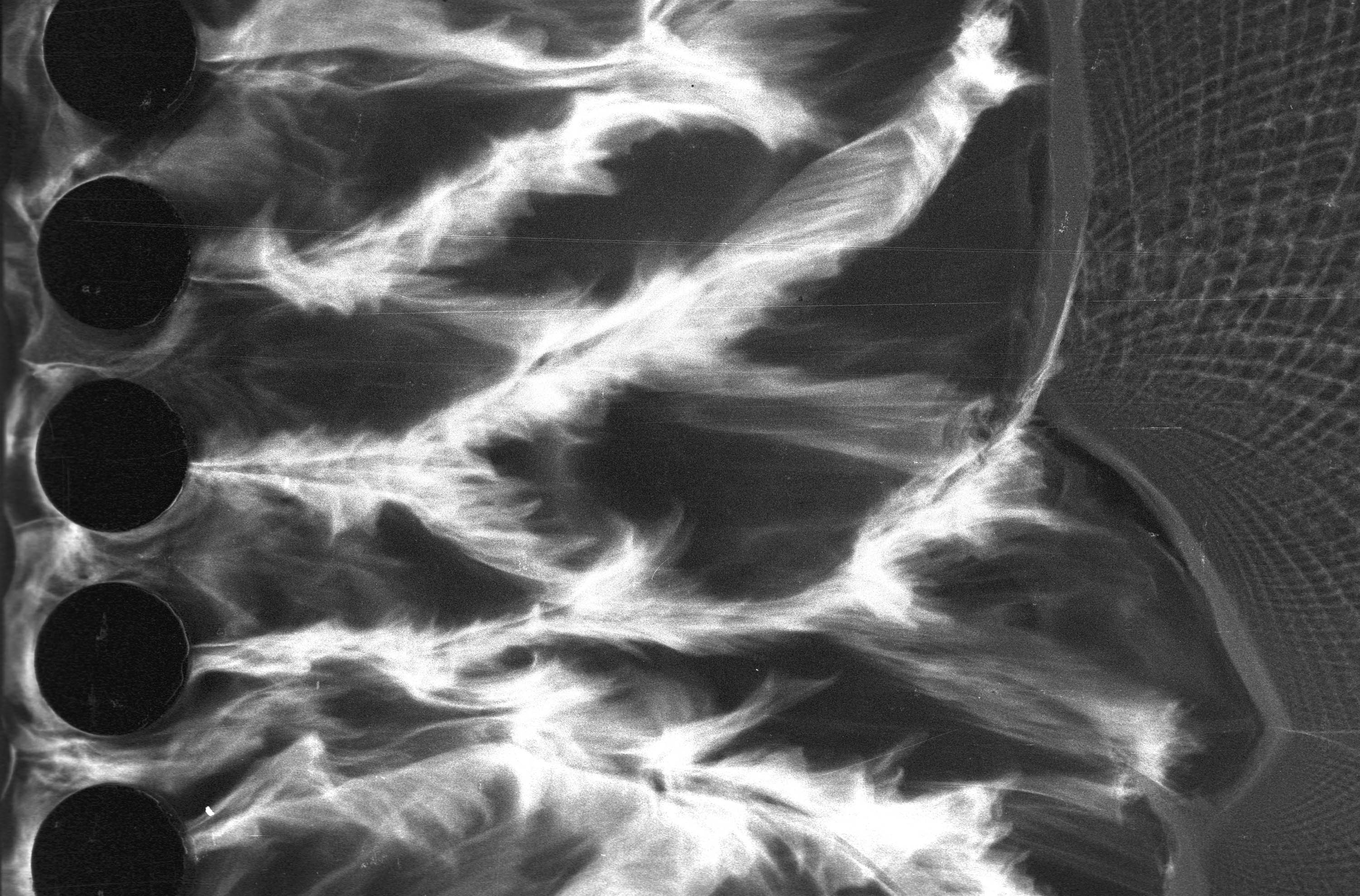}
\end{center}
\caption{Open shutter photograph illustrating the reactions along the trajectories of shear layers and the eventual onset of detonation in acetylene-oxygen, adapted from \cite{Radulescu&Maxwell2011}.}
\label{fig:openshutter}
\end{figure}

The flow field established immediately behind the obstacles (the near field of the problem) has attracted much attention in recent years.  Ohyagi et al. and Teodorczyk et et al. have focused on the conditions leading to the prompt re-initiation of the detonation from Mach reflections \cite{Obaraetal2008, Teodorczyketal1988}.  Radulescu and Maxwell \cite{Radulescu&Maxwell2011} investigated the re-initiation of detonation waves downstream of a porous medium, which comprised a two-dimensional array of staggered cylinders. They reported that such high-speed deflagrations require auto-ignition spots via shock compressions to drive strong pressure wave activity.  Their open-shutter photographs suggested the presence of intense turbulent mixing along shear layers following these punctuated hotspot ignitions, as shown in Fig. \ref{fig:openshutter}.  This suggests that not only hot-spot auto-ignitions are at play, but also turbulent mixing.  Bhattacharjee et al. focused on more delayed re-initiation in an unstable mixture of methane - oxygen, and found that the detonation re-establishment is favored by strong jets formed behind the Mach stem \cite{Bhattacharjee2013, Bhattacharjeeetal2013}. These strong internal jets and large vorticity levels are associated with the problem of Mach reflections \cite{Collela&Glaz1984, Mach&Radulescu2011, Mahmoudi&Mazaheri2011}; these jets are favored in gases with a low isentropic exponent and absent for sufficiently high values of the isentropic exponent. 

The present study focuses on the mechanism controlling the fast flame propagation in the far field, with a focus on the hydrodynamic effects.  Mixtures with different isentropic exponents but otherwise similar chemical kinetic parameters conducive to hot spot ignition are considered, in order to evaluate the importance of the hydrodynamic mechanism proposed by Bhattacharjee et al \cite{Bhattacharjee2013, Bhattacharjeeetal2013} for the near field. The experiments use the technique of detonation attenuation by single or multiple cylindrical obstacles in order generate these high speed deflagrations and observe them at different scales, hence permitting to construct a rational evolution of the flow field from well defined initial conditions in the near field to the far-field self-supported fast flames.  
  
%%%%%%%%%%%%%%%%%%%%%%%%%%%%%%%%%%%%%%%%%%%%%%%%%%%%%%%%%%%%%%%%%%%%%%%%%%
\section*{Experimental Set-up}
The experiments were conducted in a 3.5-m-long thin rectangular channel, 203-mm-tall and 19-mm-wide, as described by Bhattacharjee \cite{Bhattacharjee2013}.  The thin aspect ratio allows for quasi-two dimensional dynamics and minimizes the visual integration across the width.  The last meter of the channel was equipped with glass windows allowing flow visualization.  Two different flow vizualisation techniques were used.  A high resolution Z-type schlieren system with a 30-cm-field of view permitted to probe small scales.  It used a 1600 W continuous xenon arclamp (Newport) as light source and a high speed video camera (Phantom v1210) to record high speed videos at approximately 86 000 fps. A second visualization technique was used to monitor the experiments over the entire 1-m viewing section of the apparatus.  A large-scale Edgerton shadowgraph technique \cite{Hargather&Settles2009} was implemented using a 2-m by 2-m retro-reflective screen using the same camera and light source. 

Two different obstacle configurations with the same blockage ratio were used to allow observations of the phenomenon at different scales: a staggered assembly of five cylinders, 30.5 mm in diameter and spaced by 10 mm (shown in Fig. \ref{fig:Oxy-Hydrogen-5C-Shadow}), and a single larger cylinder, 152 mm in diameter, positioned centrally in the detonation channel. These obstacles were placed at the entrance of the visual section of the shock tube to allow for visualization of the fast flames established downstream.

We studied stoichiometric methane-oxygen and hydrogen-oxygen, which, as we will show below, have comparable chemical kinetic characteristics at the gas conditions generated by the detonation interaction with the cylindrical obstacles. The gases were mixed in a separate vessel and left to mix for a minimum of 24 hours before an experiment. Varying the initial pressure, $p_0$, of the test mixture permitted us to control the reactivity of the mixture. The channel was evacuated to below 80 Pa absolute pressure before filling with the test mixture. A high voltage spark obtained from a capacitive discharge (1000 J with deposition time of $2 \mu s$) was used to ignite the test mixtures.  In all tests, pressure gauges positioned prior to the obstacles permitted to verify that self-sustained incident detonations were established in each test.   

%%%%%%%%%%%%%%%%%%%%%%%%%%%%%%%%%%%%%%%%%%%%%%%%%%%%%%%%%%%%%%%%%%%%%%%%%%%%%%%%%%%%%%%%%%%%%%%%%
\section*{Results}
\begin{figure*}
\begin{center}
\includegraphics[width=\textwidth]{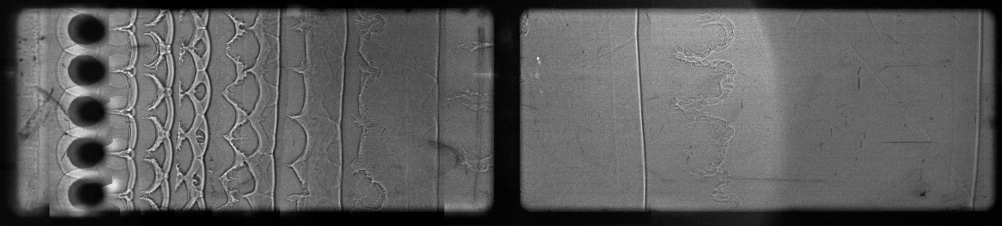}
\end{center}
\caption{Composite of shadowgraph records illustrating the fast flame decay in 2H$_2$-O$_2$, $p_0=$ 17.2 kPa; see \textit{Video 1} for the animation of sequential frames.}
\label{fig:Oxy-Hydrogen-5C-Shadow}
\end{figure*}

The maximal pressure for conducting the experiments in the transparent section of the tube was 19 kPa.  All experiments conducted at pressures below this value in hydrogen-oxygen with the five cylinders caused detonation failure.  A typical example of the flow field evolution observed in the experiments is shown in Fig.  \ref{fig:Oxy-Hydrogen-5C-Shadow} for an initial pressure of 17 kPa; the entire sequence of frames is shown in \textit{Video 1}.  For reference, these conditions correspond to $b/\lambda \simeq 1.5$, where $b$ is the gap spacing between the cylinders and $\lambda$ is the characteristic detonation cell size, taken from the Detonation Database \cite{DDB}.

Fig. \ref{fig:Oxy-Hydrogen-5C-Shadow} illustrates the initial interaction of the detonation wave and the subsequent shock-deflagration structure.  The diffraction of the incident detonation around the cylinders locally attenuates the front.  The shock and reaction zone of the detonation decouple.  The lead shock waves then reflect, causing a first generation of local hot spots, the regions of reacted gas referred to as \textit{keystone structures}. Upon the next reflection of the associated transverse waves, new Mach shocks form, which in turn give rise to a second generation of hot spots.  This sequential initiation of hot spots, without the establishment of a detonation, is consistent with the mechanism deduced by Radulescu \& Maxwell from their simulations and previous experiments in acetylene-oxygen-argon mixtures performed by Lee and Papyrin, reported in \cite{Radulescu&Maxwell2011}.  The second generation of hot spots is unable to amplify the shock waves enough to sustain them.  The transverse waves and lead shock are continuously attenuated, the reaction zone is seen to trail significantly behind the front and propagates as a flame with a bumpy structure set-up by the initial obstacle spacing.  

\begin{figure}
\centering
\includegraphics[width=0.8\textwidth]{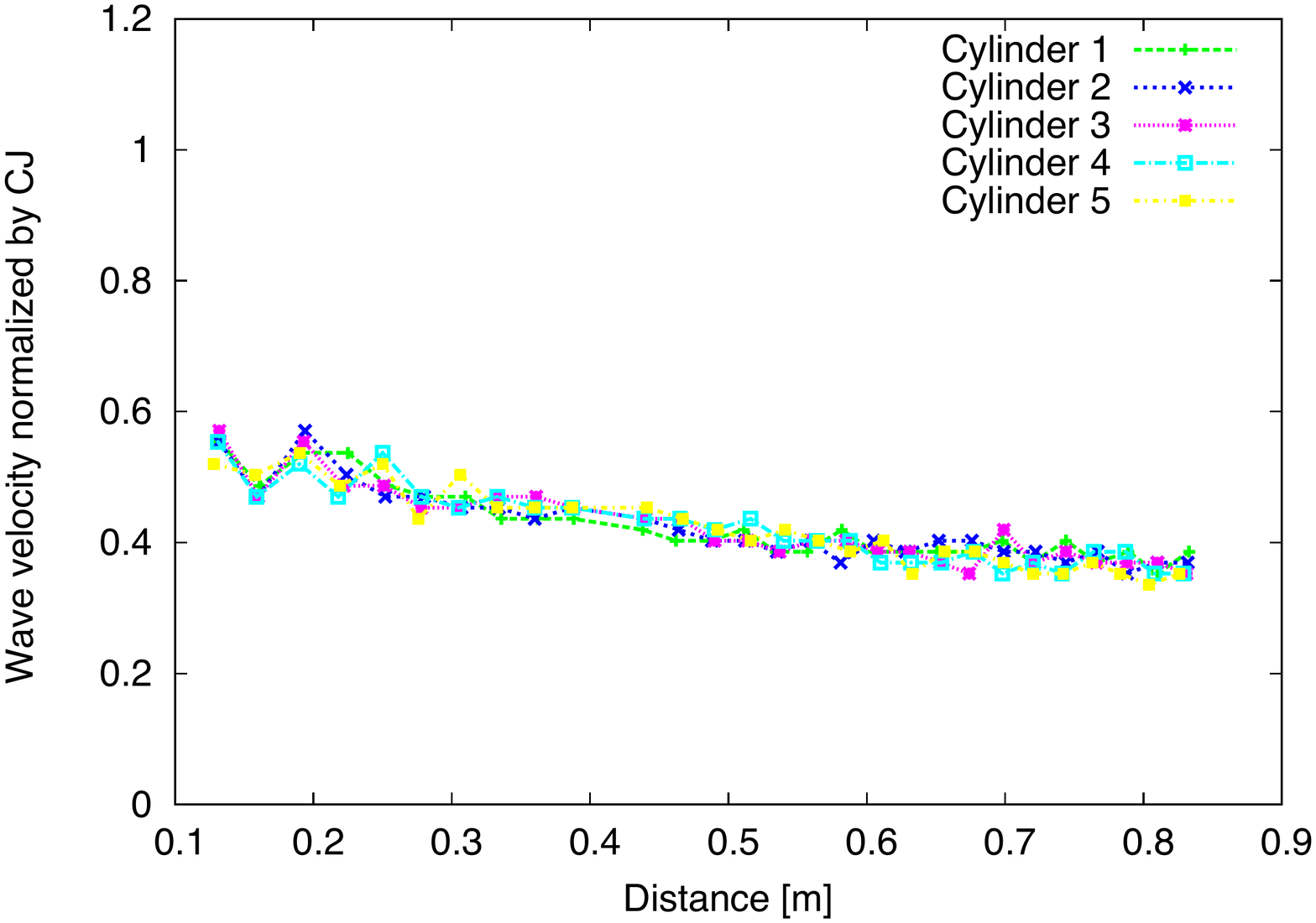}
\caption{Shock speed profiles corresponding to Fig. \ref{fig:Oxy-Hydrogen-5C-Shadow} in 2H$_2$-O$_2$.}
\label{fig:velocityHydrogen}
\end{figure}

Figure \ref{fig:velocityHydrogen} shows the lead shock speed evolution recorded along straight lines behind each of the five cylinders.  Following the obstacle interaction, the two generations of hot spots cause small increases in the shock speed before the planar lead shock starts decaying.  The shock velocity approaches that of our numerical prediction by completely turning off the chemical reactions; the numerical method used is discussed elsewhere \cite{Bhattacharjeeetal2013}.  

The experiments performed in methane-oxygen showed a significant different outcome. Self-sustained fast flames were observed over a wide range of initial conditions.  The upper limit corresponds to the criterion for prompt re-initiation behind the obstacle, which occurs at approximately at $d/\lambda \approx 2$ \cite{Bhattacharjee2013}.  Fast flames were observed for more insensitive mixtures over an order of magnitude in changes of mixture sensitivity, down to $b/\lambda \approx 0.1$, which was the limit of our experiments due to spark ignition constraints.  

\begin{figure*}
\begin{center}
\includegraphics[width=\textwidth]{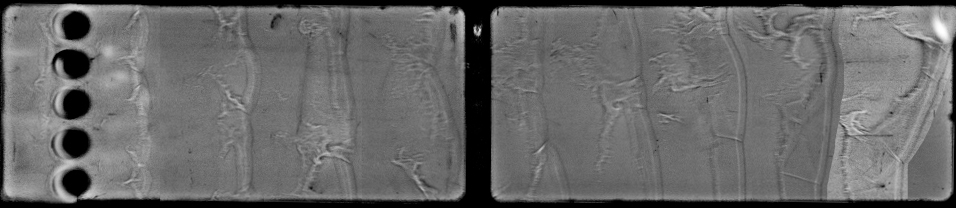}
\end{center}
\caption{Composite of shadowgraph records illustrating the fast flame sustenance in CH$_4$-2O$_2$, $p_0=$ 7.6kPa; see \textit{Video 2} for the animation of sequential frames.}
\label{fig:Oxy-Methane-5C-Shadow}
\end{figure*}

The detonation interaction at an initial pressure of 8.2 kPa, corresponding to $b/\lambda \approx 0.14$, is shown in Fig. \ref{fig:Oxy-Methane-5C-Shadow}; \textit{Video 2} shows an animation of all the sequential frames.  This experiment shows the detonation wave entering from the left, and interacting with the obstacles.  The shock diffraction at very early times is similar to the hydrogen-oxygen experiments.  However, the hot spots generated affect the shock dynamics leading to a stronger second generation of hot spots.  The second generation of hot spots is not symmetrical.  Three hot spots seem to dominate over the others and produce a complex wave structure.  This highly variable wave structure is maintained over the length of the channel, with punctuated exothermic events.  The wave structure resembles that of a highly unstable detonations \cite{Radulescuetal2005}.  The structure organizes into fewer stronger modes, as previuously observed \cite{Radulescuetal2005, Grondin&Lee2010} with an eventual transition to detonation just prior to the wave arrival at the end of the channel.  Repeat experiments showed that the phenomenon was quite reproducible on large scales. A fast meta-stable reaction wave was established in all cases with the transition to detonation near the end of the channel.  

\begin{figure}
\centering
\includegraphics[width=0.8\textwidth]{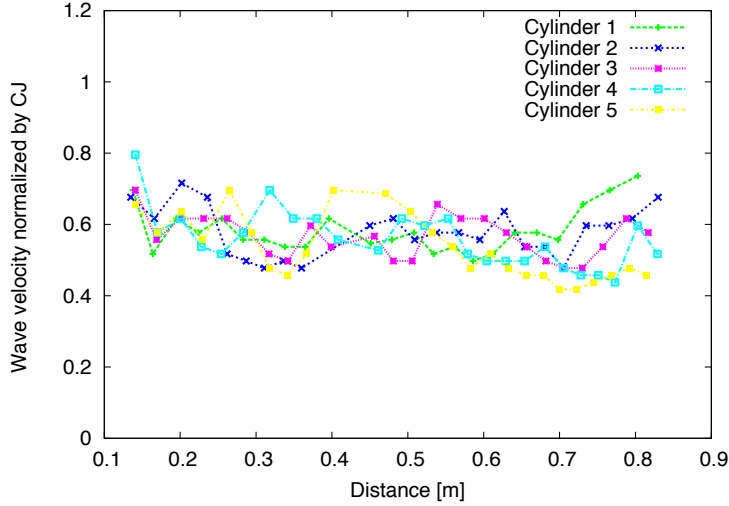}
\caption{Shock speed profiles corresponding to Fig. \ref{fig:Oxy-Methane-5C-Shadow} in CH$_4$-2O$_2$.}
\label{fig:velocityMethane}
\end{figure}

The shock speed records obtained from the analysis of sequential frames are shown in Fig. \ref{fig:velocityMethane}.  In contrast with the results obtained for hydrogen, the lead shock does not decay but is maintained quasi-steadily at nearly 0.6 of the Chapman Jouguet detonation speed.  Significant fluctuations on the order of a few 100 m/s are also observed.

\begin{figure}
\begin{center}
\includegraphics[width=0.8\textwidth]{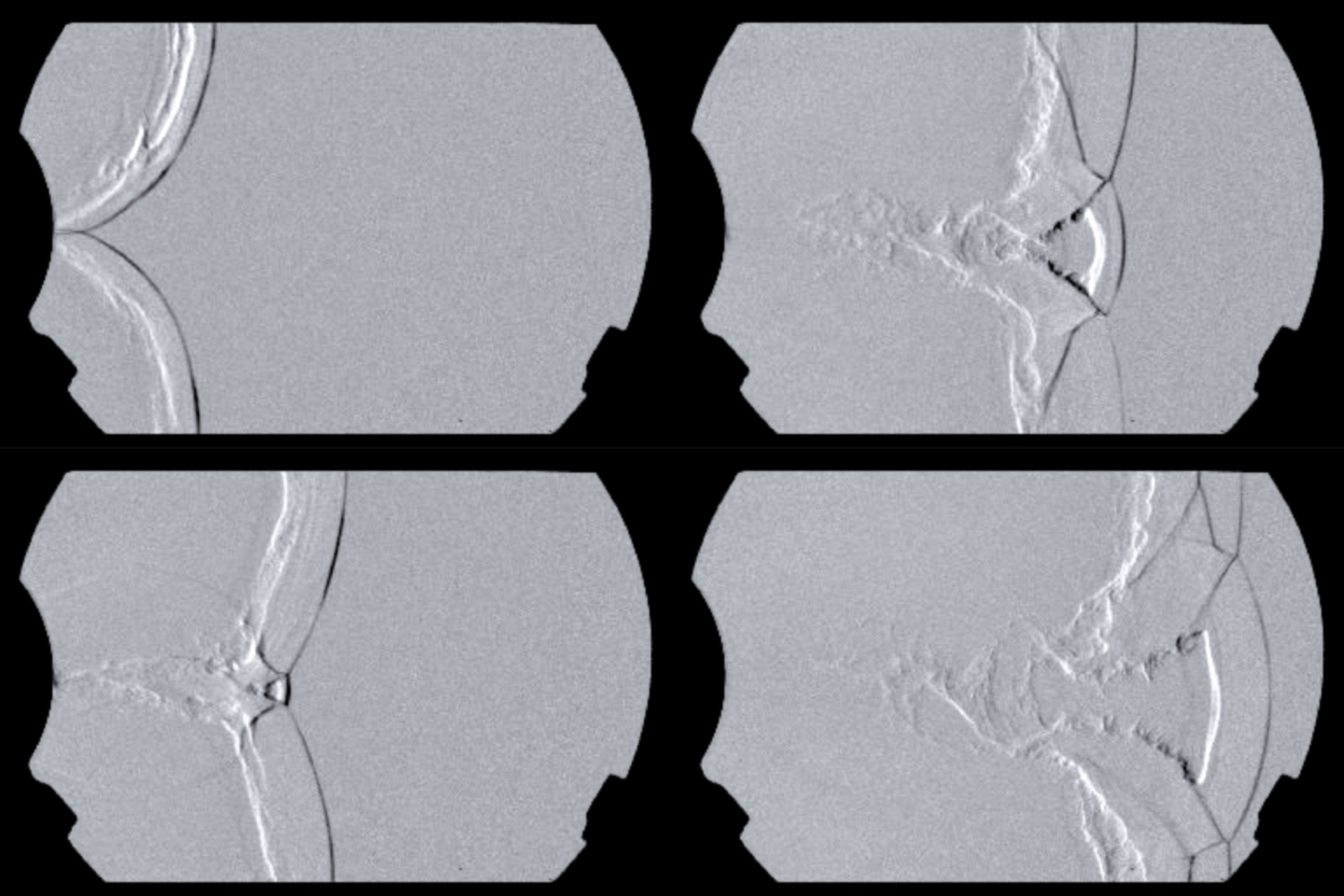}
\end{center}
\caption{Hot spot re-ignition from reflection pattern in 2H$_2$-O$_2$,$p_0$= 11.0 kPa; see \textit{Video 3} for the animation of sequential frames.}
\label{fig:Oxy-hydrogen-1C-Schlieren}
\end{figure}

To further clarify these complex wave interactions and differences between the methane and hydrogen experiments, a new set of experiments were performed with a five times (geometrically similar) larger single cylinder.  For the hydrogen mixture, the transition between prompt detonation re-initiation behind the obstacle and completely decoupled shock flame structure was at an initial pressure of approximately $p_0$ = 11 kPa, which corresponds to $b/\lambda \approx 3$.  A typical evolution of the flow field observed at these conditions is illustrated in Fig. \ref{fig:Oxy-hydrogen-1C-Schlieren}.  \textit{Video 3} shows the animation of the sequential frames obtained.   Following the interaction of the decaying lead shocks, a hot spot is first formed where the two slip lines meet; this is the gas shocked by the stronger Mach shock, that has been shocked for the longest time.  It is where the first ignition is expected.  The transient observed is a scaled-up version of the transient observed for the small cylinders illustrated in Fig. \ref{fig:Oxy-Hydrogen-5C-Shadow}. The combustion zone appears to be quite laminar, although some Kelvin-Helmholtz instability is observed on the shear layers.  

\begin{figure}
\begin{center}
\includegraphics[width=0.8\textwidth]{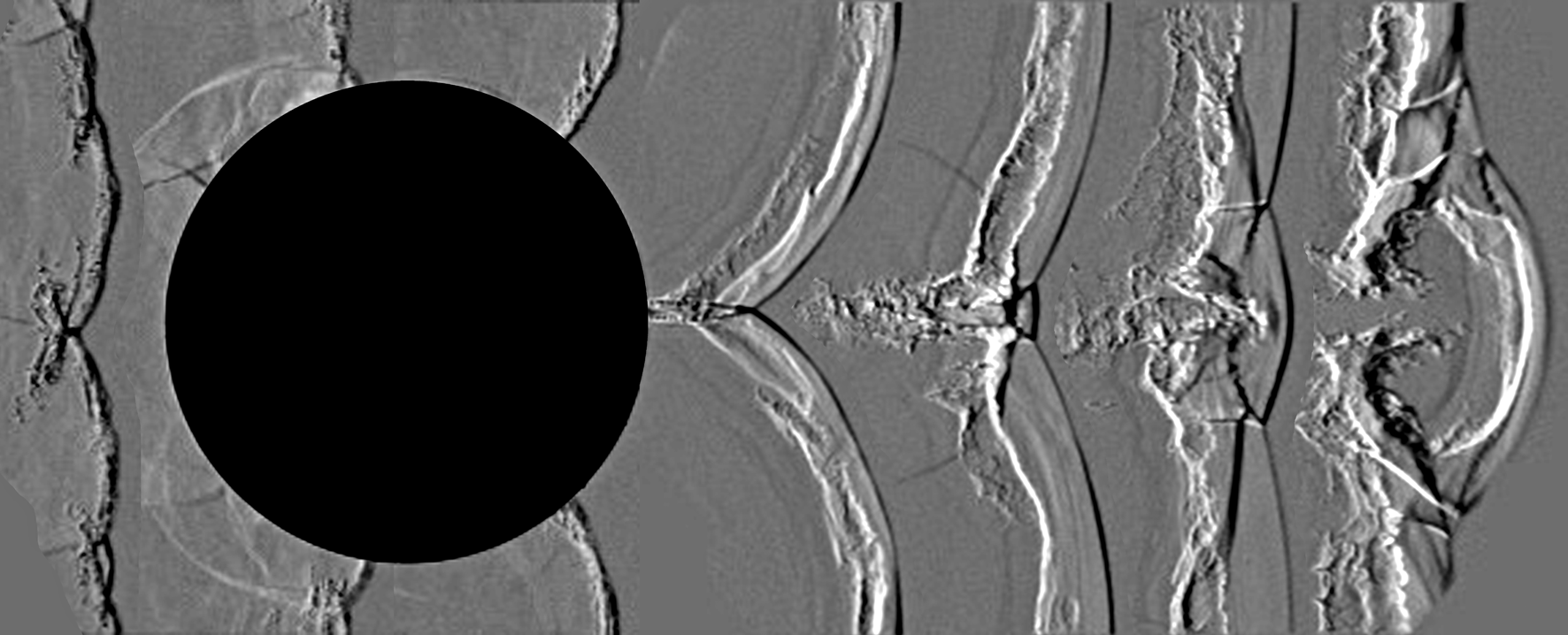}
\end{center}
\caption{Composite of Schlieren frames illustrating the diffraction of a CH$_4$+2O$_2$ detonation wave around the cylindrical obstacle and the hot-spot entrainment behind the Mach shock; $p_0$= 8.2 kPa, ; see \textit{Video 4} for the animation of sequential frames.}
\label{fig:Oxy-Methane-1C-Schlieren}
\end{figure}

A different transient is observed in the methane experiments.  A typical experiment obtained at an initial pressure $p_0$ = 8.2 kPa (corresponding to $b/\lambda = 0.8$ is shown in detail in Figure \ref{fig:Oxy-Methane-1C-Schlieren}; the evolution of the sequential frames is shown in \textit{video 4}. Figure \ref{fig:Oxy-Methane-1C-Schlieren} shows a similar hot spot ignition near the meeting point of the shear layers.  However, hydrodynamic evolution of the hot spot growth reveals that it is driven forward by a strong hydrodynamic jet.  This is consistent with the jet observed at similar conditions by Bhattacharjee et al. \cite{Bhattacharjeeetal2013}. Following this jet entrainment, the entire gas behind the Mach shock is then ignited.  Strong pressure waves ensue, with secondary triple points on the lead shock structure.  Note also the further turbulization of the shear layers from the passage of these transverse shocks.    

\begin{figure}
\centering
\includegraphics[width=0.8\textwidth]{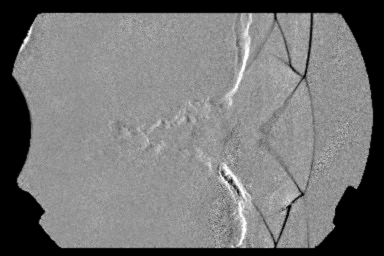}
\caption{Inert shock reflection pattern in 2H$_2$-O$_2$,$p_0$= 10.3 kPa; see \textit{Video 5} for the animation of sequential frames.}
\label{fig:InertShockRelfectionsHydrogen}
\end{figure}

\begin{figure}
\centering
\includegraphics[width=0.8\textwidth]{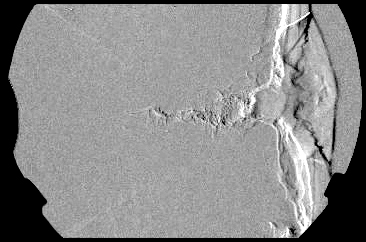}
\caption{Inert shock reflection pattern in CH$_4$+2O$_2$; $p_0$= 3.44 kPa; see \textit{Video 6} for the animation of sequential frames.}
\label{fig:InertShockRelfectionsMethane}
\end{figure}

The comparison of Figs. \ref{fig:Oxy-hydrogen-1C-Schlieren} and \ref{fig:Oxy-Methane-1C-Schlieren} shows that the dynamics of the hot spot growth is the major difference between the two cases.  To further examine these features, experiments were repeated at lower initial pressures, i.e., in less sensitive mixtures.  In sufficiently insensitive mixtures, it was found that the hot spot ignition can be eliminated on the time scales of the Mach reflection, permitting to focus on the purely hydrodynamic effects.  Figures  \ref{fig:InertShockRelfectionsHydrogen} and \ref{fig:InertShockRelfectionsMethane} show respectively the resulting inert shock reflection characteristics in hydrogen-oxygen and methane-oxygen; \textit{videos} 5 and 6 show the frame animations for each mixture, respectively. The forward jet instability behind the Mach stem was found to be much more prominent in methane-oxygen than in hydrogen-oxygen, where it was absent. Note however that the shear layers do become unstable from Kelvin Helmholtz instability in both cases. 

\section*{Discussion}
To determine the origin of the different behavior in these two mixtures, the thermo-chemical and hydrodynamic properties of each mixture were evaluated at the post-shock conditions behind the Mach shocks shown in Figures \ref{fig:Oxy-Methane-1C-Schlieren} and \ref{fig:Oxy-hydrogen-1C-Schlieren} (given the Mach number obtained from the sequential frames).  For each case the Mach stem was travelling at approximately 70\% of the Chapman-Jouguet velocity.  The post shock thermodynamic state was evaluated using frozen chemistry and real specific heats, using the NASA's Chemical Equilibrium with Applications (CEA) code.  The ignition delay, characteristic energy deposition time, and effective activation energies were calculated by conducting constant volume homogeneous reactor calculations using the CANTERA code \cite{Goodwin} and the GRI 3.0 chemical kinetic mechanism for methane \cite{GRI3} and the Li et. al. \cite{Lietal2004} mechanism for hydrogen.  The activation energy was extracted by perturbing the initial temperature in order to determine the exponential sensitivity of the ignition delay to this temperature change.  The laminar flame speed in the shocked unreacted gas was also estimated with the Cantera code.  The isentropic exponent was calculated assuming frozen chemistry at the post shock condition using the CEA code; similarly, the kinematic viscosity in the gases was estimated using the CEA code.  The velocity difference on the slip line was calculated using shock polar analysis, under the assumptions of a perfect gas.  These properties are listed in Table \ref{Tab:ChemKinetics}.

\begin{table}
\caption{Thermo-chemical properties of the gas behind the Mach shocks in hydrogen and methane}
\label{Tab:ChemKinetics}
\centering
\begin{tabular}{l c c c}
	&	&	Hydrogen-oxygen	 & Methane-oxygen				\\
\hline
%Mach shock strength & $M_S$ &  3.5 & 4.5\\
%shock state 		&  $p_s$ (kPa), $T_s$ (K) 				& 207, 980 		&209, 1105\\
reduced activation energy &$\frac{Ea}{R T_{s}}$					&37							& 17\\
ignition delay ($\mu s$)	 &$t_{i}$ 										&4.5 $\times 10^3$ &1.57$\times 10^4$ \\
reaction time	($\mu s$) &$t_{r}$ 										&0.46							& 0.47			\\
instability parameter $\chi$ & $\frac{Ea}{RT_s}\frac{t_i}{t_r}$ 	& 5.7 $\times 10^5$				& 3.6 $\times 10^5$\\
isentropic exponent 	& $\gamma$  								&1.36  							&1.22\\
laminar flame speed (m/s) 	&$S_{L}$	 							&64							&17	\\
kinematic viscosity (m$^2$/s) &  $\nu$ & 1.4 $\times 10^{-4}$ & 7.8 $\times 10^{-5}$\\
Velocity difference on slip line (m/s) &  & 581 & 512\\
\hline
\end{tabular}
\end{table}

In the past, it has been suggested that fast flames can be correlated with the propensity to form hot spots \cite{Kuznetsovetal2000, Chao2006, Grondin&Lee2010, Radulescuetal2013}, as discussed in the Introduction.  This would be reflected by large reduced activation energies and high $\chi$ values \cite{Liangetal2007, Radulescuetal2013, Bradley2012}.  Table \ref{Tab:ChemKinetics} shows that for these cases the activation energy of the shocked hydrogen-oxygen gas is higher than that of methane-oxygen at the relevant thermodynamic state behind the Mach shocks of the fast flame. Likewise, the values of $\chi$ are comparable at these conditions.  Although these parameters indicate that hotspots with strong pressure generation will form in both mixtures \cite{Bradley2012, Radulescuetal2013}, we cannot discriminate between the two mixtures.  This suggests that the fundamental differences observed in our experiments between the two mixtures cannot be reconciled by thermo-kinetic arguments relating the propensity to form hot spots.  
  
It is also expected that a mixture with a shorter ignition delay is more likely to ignite.  The conditions behind the Mach stem indicate that hydrogen would be more likely to ignite due to the shorter predicted ignition delay. This again cannot account for the experimental observations.  

The calculated laminar flame speed in oxy-hydrogen is higher than that of oxy-methane.  This indicates that laminar heat diffusion is not the dominating propagation mechanism.  

\begin{figure}
\centering
\includegraphics[width=0.8\textwidth]{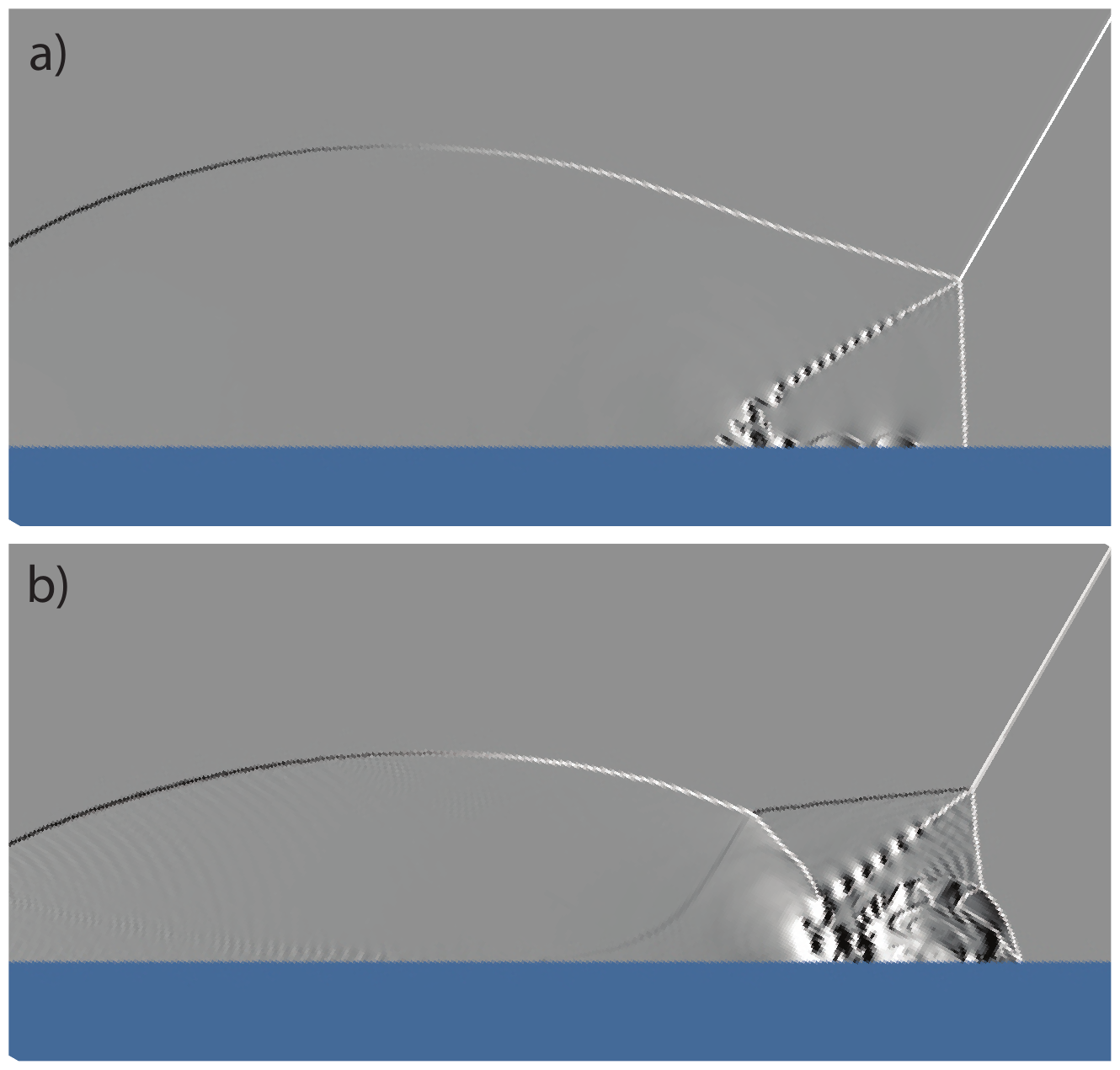}
\caption{Inert shock reflections in a model hydrogen-oxygen and methane-oxygen mixtures; the incident shock Mach number is taken as $1/2M_{CJ}$ for both cases, 30 $^\circ$ incidence angle and $\gamma = 1.36$ and 1.2 respectively.}
\label{fig:Simulations-Reflect}
\end{figure}

We now turn to the hydrodynamic properties of the mixtures.  The specific heat ratio of the oxy-methane mixture is significantly lower than that of the oxy-hydrogen mixture.  Recently, Mach and Radulescu have shown that for mixtures with lower specific heat ratios more prominent forward jetting of the slip line will would occur\cite{Mach&Radulescu2011} in Mach reflections.  This is in good agreement with the slip lines seen in Figures \ref{fig:InertShockRelfectionsMethane} and \ref{fig:InertShockRelfectionsHydrogen}.  To further substantiate this mechanism, we have calculated the shock reflection problem in an inert gas using the perfect gas approximation using the incident shock strength, reflection angle and isentropic exponents of the experiments by the method outlined in \cite{Mach&Radulescu2011}.  Figure  \ref{fig:Simulations-Reflect} shows the reflection patterns expected in the hydrogen and methane gases, respectively.  While in the hydrogen mixture, a very weak jet is driven forward during the reflection, a much stronger jet is evident for the methane-oxygen system, characterized by a recirculation region and disturbances of the lead Mach shock.  

In summary, both the experiments and theoretical arguments suggest that hydrodynamic instabilities (strong vorticity) associated with shock reflections are the mechanisms to enhance the mixing in the reaction zone structure of high speed deflagrations.  The importance of the isentropic exponent in this respect is to recognize that vorticity in compressible media is driven primarily by the baroclinic torque mechanism of misaligned density and pressure gradients.  Lowering the isentropic exponent gives rise to a more compressible medium, with correspondingly larger density gradients.  This leads to higher rates of vorticity production. 

\begin{figure}
\begin{center}
\includegraphics[width=0.8\textwidth]{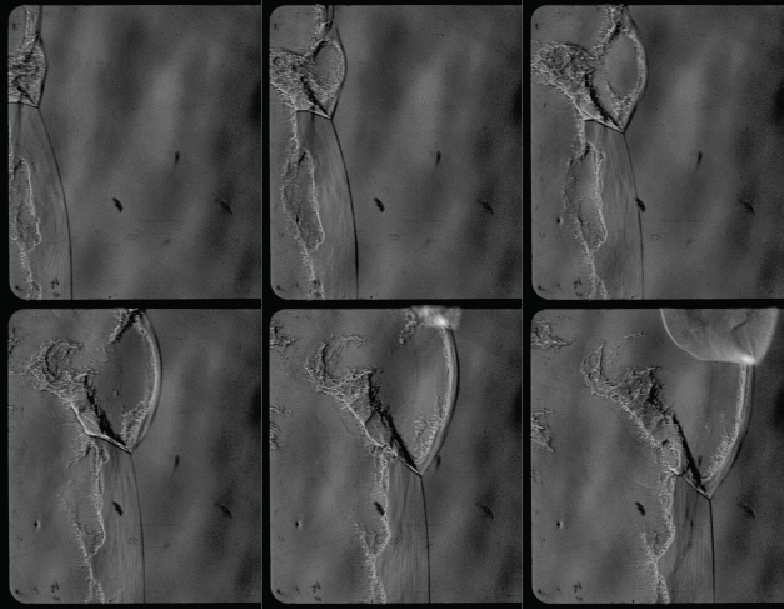}
\end{center}
\caption{Sequential frames illustrating the local hot-spot ignition in the structure of fast flames, jet entrainment and local explosion driving pressure waves, with eventual onset of detonation in CH$_4$+2O$_2$; images were taken approximately 3 cylinder diamaters downstream of the 50-mm-diameter cylinder $p_0$= 8.4 kPa; see \textit{Video 7} for the animation of sequential frames.}
\label{fig:reflect-ignite}
\end{figure}

Analysis of the high speed movies obtained permitted us to infer the mechanism by which an increase in vorticity may lead to the enhancement of the burning rate and sustenance of a high speed regime and the eventual acceleration to detonation.  Figure \ref{fig:reflect-ignite} shows an example of this behavior in methane-oxygen, observed in the single cylinder experiments with the field of view approximately three cylinder diameters downstream of the obstacle.  \textit{Video 7} shows the evolution of the local flowfield.  Local shock reflections give rise to local hydrodynamic jets, rapid mixing and fast combustion rates. The resulting reaction appears to be sufficiently fast to trigger its own system of pressure waves.  Existing modes thus further amplify until one of them becomes sufficiently strong to trigger local detonative combustion. 

It is of interest to assess the intense mode of turbulent combustion in the jet entrainment apparent in the experiments (e.g., Fig. \ref{fig:Oxy-Methane-1C-Schlieren}).  Using standard Kolmogorov scaling arguments relating the integral scale quantities (denoted by subscript $i$) to the Kolmogorov scales (subscript $\eta$), it can be shown that the characteristic eddy turn-over time at the Kolmogorov scales (the ones with shortest time scale) is given by $t_{\eta}=\sqrt{\nu l_i/V_i^3}$.  Taking the integral scale as $l_i\simeq2$ cm apparent from the records, the characteristic shear on the slip line calculated (see Table \ref{Tab:ChemKinetics}) as characteristic shear $V_i$, the Kolmogorov time becomes approximately $t{_\eta}\simeq 10^{-1} \mu$sec.  This time scale is substantially smaller than the characteristic flame passage time, which is given by the characteristic energy deposition time $t_r$.  This means that the burning regime in the coherent jets forming these high speed deflagrations is likely in the distributed flame regime.  This is consistent with our experiments, which suggest that high burning rates are established in these coherent structures, sufficiently large to yield noticeable pressure evolution.  This appears to be the burning mechanism of high speed turbulent deflagrations.  

\section*{Conclusions}

The present experiments suggest that one factor controlling the propensity of a reactive mixture to support fast flames is its ability to generate vorticity by hydrodynamic instabilities.  These appear to be favored in mixtures with a lower isentropic exponent, for which density changes are more significant.  The experiments have shown that such hydrodynamic instabilities may couple with the reactive field when auto-ignition is possible in discrete regions.  The punctuated reaction of these spots permits to drive the pressure fluctuations and shock amplitude changes conducive to more instabilities.  

While these implications were made possible by having chosen two reactive systems that share similar thermo-kinetic parameters governing the rate at which energy is released in local hotspots (hence ruling out these effects from consideration), future work should be devoted at the coupling between these two aspects of the problem.  Particularly, it would be of interest to probe two mixtures that are equally prone to hydrodynamic instabilities but for which the sensitivities to create hot spots are significantly different.

%%%%%%%%%%%%%%%%%%%%%%%%%%%%%%%%%%%%%%%%%%%%%%%%%%%%%%%%%%%%
\section*{Acknowledgements}
We wish to thank NSERC for financial support through a Discovery Grant to M.I.R. and the H2CAN Strategic Network of Excellence.   M.I.R dedicates this paper to Ugo Bugo and acknowledges useful discussions with John Lee, Alexei Poludnenko and Joe Shepherd.
%% References


%\bibliographystyle{elsarticle-num-CNF}
\bibliographystyle{ieeetr}
\bibliography{references}






% \end{thebibliography}

\end{document}